\documentclass[prd,onecolumn,nofootinbib,showpacs,showkeys]{revtex4}
\usepackage{bm,amsmath,amssymb}
\newcommand\gothfamily{\usefont{U}{ygoth}{m}{n}}
\DeclareTextFontCommand{\textgoth}{\gothfamily}

\begin{document}

\title{Conformal time in a black-hole universe with torsion}
\author{Nikodem J. Pop{\l}awski}
\affiliation{Department of Physics, Indiana University, Swain Hall West, 727 East Third Street, Bloomington, Indiana 47405, USA}
\email{nipoplaw@indiana.edu}
\date{\today}

\begin{abstract}
In the Einstein-Cartan-Sciama-Kibble theory of gravity, the intrinsic spin of fermionic matter generates spacetime torsion and induces gravitational repulsion at extremely high densities.
This repulsion prevents the collapsing spin-fluid matter in a black hole from forming a singularity.
Instead, the interior of a black hole with a stiff equation of state becomes a new universe, which contracts until a (big) bounce and then expands.
We derive the equations describing the dynamics of our Universe, formed in such a scenario, in terms of the conformal time which is a convenient variable for testing signatures of the contracting phase in the Cosmic Microwave Background radiation.
\end{abstract}

\maketitle

The Einstein-Cartan-Sciama-Kibble (ECSK) theory of gravity removes the constraint of general relativity that the torsion tensor (the antisymmetric part of the affine connection) be zero by promoting this tensor to a dynamical variable \cite{KS,Hehl}.
In this theory, which has no free parameters, the Lagrangian density for the gravitational field is proportional to the curvature scalar.
The field equations follow from the stationarity of the total action under the variations of the metric and torsion tensors.
These equations naturally account for the quantum-mechanical, intrinsic angular momentum (spin) of elementary particles.
Since Dirac fields minimally couple to the affine connection, the torsion tensor in fermionic matter that composes all stars in the Universe is different from zero.
Macroscopic averaging of the conservation law for this spin tensor leads to the description of fermionic matter as a spin fluid \cite{NSH}.

The field equations of the ECSK gravity for a spin fluid can be written as the general-relativistic Einstein equations with the modified energy-momentum tensor that has terms which are quadratic in the spin tensor \cite{KS,Hehl}.
These terms generate gravitational repulsion in fermionic matter which is significant only at densities that are much larger than the nuclear density, that is, in the early Universe and inside black holes.
This repulsion prevents the formation of unphysical singularities \cite{avert,HHK} and replaces the singular big bang by a nonsingular bounce, before which the Universe was contracting \cite{Kuch,bounce,infl}.
Such a bounce naturally explains why the present Universe appears spatially flat, homogeneous, and isotropic, without needing cosmic inflation \cite{infl}.
Torsion also modifies the classical Dirac equation by generating the cubic Hehl-Datta term \cite{Dirac}, which may be the source of the observed matter-antimatter imbalance and dark matter in the Universe \cite{bar}.
Furthermore, the gravitational interaction of condensing fermions due to the Hehl-Datta term may be the source of dark energy \cite{dark}.
In this paper, we describe the dynamics of the Universe with a torsion-induced big bounce in terms of the conformal time.

We consider a closed ($k=1$), homogeneous, and isotropic universe filled with fermionic matter macroscopically averaged as a spin fluid \cite{HHK,Kuch}.
The Einstein-Cartan field equations for the Friedman-Lema\^{i}tre-Robertson-Walker (FLRW) metric describing such a universe are given by the Friedman equations for the scale factor $a(t)$ \cite{infl}:
\begin{eqnarray}
& & {\dot{a}}^2+k=\frac{1}{3}\kappa\Bigl(\epsilon-\frac{1}{32}\kappa(\hbar cn)^2\Bigr)a^2+\frac{1}{3}\Lambda a^2, \label{Fri1} \\
& & \frac{d}{dt}\bigl((\epsilon-\kappa(\hbar cn)^2/32)a^3\bigr)+(p-\kappa(\hbar cn)^2/32)\frac{d}{dt}(a^3)=0,
\label{Fri2}
\end{eqnarray}
where dot denotes differentiation with respect to $ct$, $n$ is the fermion number density, and $\Lambda$ is the cosmological constant.
The spin-density contribution to the total energy density in (\ref{Fri1}) scales according to \cite{Kuch,infl}
\begin{equation}
\epsilon_S=-\frac{1}{32}\kappa(\hbar cn)^2\propto a^{-6},
\end{equation}
whereas the other components (indexed by $i$) scale according to $\epsilon_i\propto a^{-3(1+w_i)}$, where $w_i=p_i/\epsilon_i$ is the barotropic pressure-to-energy density ratio.
The energy density of a spin fluid in the Universe is thus
\begin{equation}
\epsilon+\epsilon_S=\epsilon_{R0}\hat{a}^{-4}+\epsilon_{M0}\hat{a}^{-3}+\epsilon_\Lambda+\epsilon_{S0}\hat{a}^{-6},
\label{tot}
\end{equation}
where $\hat{a}=a/a_0$ is the normalized scale factor, subscripts $0$ denote quantities measured at the present time (when $\hat{a}=1$), $R$ denotes radiation, $M$ denotes nonrelativistic matter (dark plus baryonic), and $\epsilon_\Lambda=\Lambda/\kappa$.
The first Friedman equation (\ref{Fri1}) can be written as \cite{infl}
\begin{equation}
|H|=H_0\bigl(\Omega_S\hat{a}^{-6}+\Omega_R\hat{a}^{-4}+\Omega_M\hat{a}^{-3}+\Omega_\Lambda-(\Omega-1)\hat{a}^{-2}\bigr)^{1/2},
\label{Hub}
\end{equation}
where $H=c\dot{a}/a$ is the Hubble parameter, $\Omega_i=\epsilon_i/\epsilon_c$ is the present-day density parameter for the $i$-component of the spin fluid, $\epsilon_c=3H^2_0/(\kappa c^2)$ is the present-day critical energy density, and $\Omega=\sum_i\Omega_i>1$ is the present-day total density parameter which satisfies $a_0 H_0\sqrt{\Omega-1}=c$.

If we introduce the conformal time $\eta$ \cite{LL},
\begin{equation}
d\eta=\frac{c\,dt}{a(t)},
\end{equation}
then (\ref{Hub}) becomes an equation for $\hat{a}(\eta)$:
\begin{equation}
\frac{c}{a_0\hat{a}^2}\biggl|\frac{d\hat{a}}{d\eta}\biggr|=H_0\bigl(\Omega_S\hat{a}^{-6}+\Omega_R\hat{a}^{-4}+\Omega_M\hat{a}^{-3}+\Omega_\Lambda-(\Omega-1)\hat{a}^{-2}\bigr)^{1/2}.
\end{equation}
This equation is equivalent to
\begin{equation}
\frac{d\hat{a}}{(\Omega_S\hat{a}^{-2}+\Omega_R+\Omega_M\hat{a}+(1-\Omega)\hat{a}^2+\Omega_\Lambda\hat{a}^4)^{1/2}}=\frac{d\eta}{(\Omega-1)^{1/2}}.
\label{conformal}
\end{equation}
The 7-year Wilkinson Microwave Anisotropy Probe (WMAP) data give $\Omega=1.002$, $H^{-1}_0=4.4\times10^{17}\,\mbox{s}$, $\Omega_R=8.8\times10^{-5}$, $\Omega_M=0.27$, and $\Omega_\Lambda=0.73$ \cite{WMAP}.
The number density of relic background neutrinos, which are the most abundant fermions in the Universe, gives an extremely small in magnitude value of the spin-torsion density parameter, $\Omega_S\approx-8.6\times10^{-70}$ \cite{infl}.

In the early Universe, which is dominated by relativistic particles, we can neglect the contributions to (\ref{Hub}) from nonrelativistic matter, the cosmological constant and the curvature ($-(\Omega-1)\hat{a}^{-2}$) because they are much smaller than $\Omega_R\hat{a}^{-4}$ (due to $\hat{a}\ll 1$).
Accordingly, (\ref{Hub}) reduces to
\begin{equation}
|H|=H_0\bigl(\Omega_S\hat{a}^{-6}+\Omega_R\hat{a}^{-4}\bigr)^{1/2}.
\label{early}
\end{equation}
Since $\Omega_S<0$, the spin-torsion coupling generates gravitational repulsion which is significant at very small $\hat{a}$, preventing the cosmological singularity at $\hat{a}=0$.
The expansion of the Universe started when $H=0$, at which the normalized scale factor had a minimum but finite value $\hat{a}=\hat{a}_m$:
\begin{equation}
\hat{a}_m=\biggl(-\frac{\Omega_S}{\Omega_R}\biggr)^{1/2}.
\label{mini}
\end{equation}
Before reaching its minimum size, the Universe was contracting with $H<0$.
In the early Universe, (\ref{conformal}) reduces to
\begin{equation}
\frac{d\hat{a}}{(\Omega_S\hat{a}^{-2}+\Omega_R)^{1/2}}=\frac{d\eta}{(\Omega-1)^{1/2}}.
\label{confo}
\end{equation}
If we choose $\eta=0$ at the minimum scale factor, $\hat{a}(0)=\hat{a}_m$, then integrating (\ref{confo}) gives
\begin{equation}
\hat{a}(\eta)=\hat{a}(0)\sqrt{1+\frac{\eta^2}{\eta^2_S}},
\label{bou}
\end{equation}
where
\begin{equation}
\eta_S=\frac{\sqrt{\Omega_S(1-\Omega)}}{\Omega_R}
\end{equation}
is the characteristic conformal time at which radiation begins to dominate over the spin-torsion coupling.
An equation of form (\ref{bou}) represents a two-component fluid composed of radiation and stiff matter, and has been considered in \cite{pert}.
The WMAP data give the characteristic conformal time at the end of the spin-torsion-dominated epoch equal to
\begin{equation}
\eta_S\approx 1.5\times 10^{-32}.
\end{equation}

As the Universe expands, $\eta\gg\eta_S$, so that (\ref{bou}) reduces to
\begin{equation}
\hat{a}(\eta)=\hat{a}(0)\frac{\eta}{\eta_S}=\sqrt{\frac{\Omega_R}{\Omega-1}}\eta.
\label{rad}
\end{equation}
As the Universe expands further, we must use (\ref{Hub}) and (\ref{conformal}) without neglecting the contributions from nonrelativistic matter, the cosmological constant and the curvature (but we can neglect the spin-density component).
The Universe becomes dominated by nonrelativistic matter, for which (\ref{conformal}) gives $d\sqrt{\hat{a}}\approx(\sqrt{\Omega_M}d\eta)/(2\sqrt{\Omega-1})$, and later by the cosmological constant, for which (\ref{conformal}) gives $d(1/\hat{a})\approx-\sqrt{\Omega_\Lambda}d\eta/\sqrt{\Omega-1}$.
The conformal time at the future infinity is equal to
\begin{equation}
\eta_\infty=\int_{\hat{a}_m}^\infty\frac{\sqrt{\Omega-1}d\hat{a}}{\sqrt{\Omega_S\hat{a}^{-2}+\Omega_R+\Omega_M\hat{a}+(1-\Omega)\hat{a}^2+\Omega_\Lambda\hat{a}^4}}.
\end{equation}
The present-day conformal time is
\begin{equation}
\eta_0=\int_{\hat{a}_m}^1\frac{\sqrt{\Omega-1}d\hat{a}}{\sqrt{\Omega_S\hat{a}^{-2}+\Omega_R+\Omega_M\hat{a}+(1-\Omega)\hat{a}^2+\Omega_\Lambda\hat{a}^4}}.
\end{equation}
The conformal time at recombination (with redshift $z_r=1089$ \cite{WMAP}) is equal to
\begin{equation}
\eta_r=\int_{\hat{a}_m}^{1/(1+z_r)}\frac{\sqrt{\Omega-1}d\hat{a}}{\sqrt{\Omega_S\hat{a}^{-2}+\Omega_R+\Omega_M\hat{a}+(1-\Omega)\hat{a}^2+\Omega_\Lambda\hat{a}^4}}.
\end{equation}
The WMAP data give
\begin{equation}
\eta_r=0.003,\,\,\,\eta_0=0.15,\,\,\,\eta_\infty=0.2.
\end{equation}

According to (\ref{Hub}), the contraction of the Universe before the big bounce looks like the time reversal of the expansion after the bounce.
A scenario in which a universe contracts from infinity in the past does not explain, however, what caused such a contraction, as the big-bang cosmology cannot explain what happened before the big bang.
It is possible, however, that the matter in the Universe before the big bounce has a different equation of state than the matter after the bounce (relativistic spin fluid).
Our Universe may thus have contracted (until the bounce) from a finite initial state such as a black hole that has formed in another universe \cite{BH}.
We showed in \cite{mass} that gravitational collapse of fermionic spin-fluid matter with a stiff equation of state, $w=1$ \cite{stiff_mat}, in a black hole of mass $M$ leads to a bounce and forms a universe of mass at least $\sim M_\ast=M^2 m_n/m_\textrm{Pl}^2$, where $m_n$ is the mass of a neutron and $m_\textrm{Pl}$ is the Planck mass.
This scenario would also explain what happens inside a black hole upon reaching the bounce: the interior of every black hole becomes a new universe \cite{infl}.
Such a scenario agrees with proposals that the matter in the early Universe has a stiff equation of state \cite{stiff_cosm} and with observations that neutron stars are composed of stiff matter \cite{stiff_NS}.

A closed universe born in a Schwarzschild black hole begins to contract when $\dot{a}=0$ at $a_i=r_g$, where $r_g=2GM/c^2$.
If the matter in a black hole is stiff then the Friedman equation describing the contracting phase of such a universe is given by \cite{mass}
\begin{equation}
{\dot{a}}^2+k=\frac{1}{3}\kappa\epsilon_i\frac{a_i^6}{a^4}-\frac{1}{96}(\hbar c\kappa)^2 n_i^2\frac{a_i^{12}}{a^{10}},
\label{inter}
\end{equation}
where $\epsilon_i$ and $n_i$ correspond to the universe at $a=a_i$ (we neglect the cosmological constant because it is negligibly small for $a\le a_i$), and $k=1$.
At $a=a_i$, the second term on the right-hand side of (\ref{master}) is also negligible, yielding $\epsilon_0=(3Mc^2)/(4\pi r_g^3)$ and $n_0=(9M)/(4\pi r^3_g m_n)$ (assuming that the matter in a black hole is composed of neutrons).
The equation (\ref{inter}) can thus be written as
\begin{equation}
{\dot{a}}^2+k=\frac{r^4_g}{a^4}-\frac{27\lambda^2_n}{128\pi^2}\frac{r^8_g}{a^{10}},
\label{master}
\end{equation}
where $\lambda_n=h/(m_n c)$ is the Compton wavelength of a neutron.
The universe in the black hole contracts until $a=a_f$, at which the negative second term on the right-hand side of (\ref{master}) counters the first term ($k$ is negligible in this regime).
The condition $\dot{a}=0$ in (\ref{master}) gives $a_f=((27r^4_g\lambda^2_n)/(128\pi^2))^{1/6}$.
Since the total mass of the stiff matter in the universe $m$ scales according to $m\propto a^{-3}$, the mass of the universe at $a=a_f$ is on the order of $M_\ast=M^2 m_n/m_\textrm{Pl}^2$ \cite{mass}.
The mass increase of a stiff spin fluid is physically realized by pair production in the presence of strong gravitational fields.

The density of matter at $a=a_f$ is on the order of $M_\ast/a^3_f$, which is on the order of the Cartan density for nonrelativistic neutrons, $\rho_{Cn}=(m^2_n c^4)/(G\hbar^2)$ \cite{non}.
Such a density is much lower than the Planck density, which is the order of the density of relativistic fermionic matter at the big bounce generated by the ECSK spin-torsion coupling \cite{bb}.
Particle-antiparticle annihilation, however, lowers the rate at which the number density of fermions (and thus the spin density) increases during the contracting phase.
A lower spin density delays reaching the regime when the gravitational repulsion due to the spin-torsion coupling stops the collapse.
The universe in a black hole therefore contracts to a scale factor lower than $a_f$ and reaches a mass higher than $M_\ast$.
This annihilation changes a stiff spin fluid into a relativistic spin fluid and allows a corrected value of $a_f$ to match $\hat{a}_m a_0$.
It also allows a corrected value of $M_\ast$ to match the mass of nonrelativistic matter in the Universe, $2\pi^2 a^3_0\Omega_M\epsilon_c/c^2$.
As a result, the contracting phase of the universe born in a black hole can be smoothly extended, through a bounce, into the expanding phase of our Universe.

In terms of the conformal time, (\ref{master}) with the second term on the right-hand side neglected becomes
\begin{equation}
\frac{da}{a\sqrt{\frac{r^4_g}{a^4}-1}}=d\eta.
\end{equation}
Integrating this equation from $a=a_i$ to $a=a_f$, at which we set $\eta=0$, gives
\begin{equation}
\eta_i\approx-\frac{\pi}{4}.
\end{equation}
Including particle-antiparticle annihilation and the relativistic part of the contraction does not change much the value of $\eta_i$.
If our Universe was born in a black hole, it formed with the black hole's event horizon.
It was contracting from $a=a_i$ (at which $\eta=-\pi/4$) to a bounce where $\eta=0$, after which it expands to infinity with a finite $\eta_\infty$.

It has been recently proposed that the Cosmic Microwave Background radiation exhibits circles of anomalously low temperature variance that may be signatures of the contracting phase of our Universe preceding the big bounce \cite{pre}.
That proposal has been criticized and currently is under debate \cite{against}.
If these circles exist, however, then the value $\theta$ such that
\begin{equation}
\mbox{sin}\,\theta=\frac{\eta_r-\eta_i}{\eta_0-\eta_r}
\end{equation}
represents the maximum angular radius of such a circle \cite{for}.
The value $\eta_i=-\pi/4$ indicates that $\mbox{sin}>1$, so that a scenario in which the Universe was born in a black hole does not put any limits on the angular sizes of such circles \cite{for}.
This prediction of the ECSK big-bounce cosmology is thus the same as that of cosmic inflation.

\end{document}